\documentclass[aps,prd,nofootinbib,showpacs,amsmath]{revtex4-1}
\usepackage{graphicx,epsf,amssymb,url,hyperref}
\usepackage{mathrsfs}
\usepackage[]{latexsym}
\usepackage[UKenglish]{babel}
\usepackage{epsfig}
\usepackage[utf8]{inputenc}

\newcommand{\be}{\begin{eqnarray}}
\newcommand{\ee}{\end{eqnarray}}

\def\comment#1{}

\usepackage{color}

\definecolor{darkred}{rgb}{.8,0,0}

\definecolor{darkblue}{rgb}{0,0,.7}

\definecolor{darkgreen}{rgb}{0,.7,0}

\begin{document}

%
%
\title{Quantum Fluctuations from Thermal Fluctuations in Jacobson Formalism} 
%

%
%
%
%
\author{Mir Faizal$^1,^2$}\email[email:~]{mirfaizalmir@googlemail.com}
\author{Amani Ashour$^3$}
\author{ Mohammad Alcheikh$^{3}$} 
\author{Lina Alasfar$^4$}\email[email:~]{lina.alasfar@outlook.fr}
\author{Salwa Alsaleh$^5$}\email[email:~]{salwams@ksu.edu.sa}
\author{ Ahmed Mahroussah$^5$}

\affiliation{$^1$Irving K. Barber School of Arts and Sciences, \\ University of British Columbia - Okanagan,\\
	Kelowna,   BC V1V 1V7, Canada }
\affiliation{$^2$Department of Physics and Astronomy,\\ University of Lethbridge,\\  Lethbridge, AB T1K 3M4, Canada}
\affiliation{$^3$Mathematics  Department,  Faculty  of  Science, \\  Damascus  university,  Damascus,  Syria }
\affiliation{$^4$ Universit\'{e} Clermont Auvergne, 4, Avenue Blaise Pascal 	63178 Aubi\`{e}re Cedex, France }
\affiliation{$^5$Department of Physics and Astronomy, King Saud University, Riyadh 11451, Saudi Arabia}

\date{\today}

	\begin{abstract}
		In the Jacobson formalism general relativity is obtained from thermodynamics. 
		This is done by using the Bekenstein-Hawking entropy-area relation.
		However, as a black holes will gets smaller, its temperature will increase. 
		This will cause the thermal fluctuations to also increase,
		and these will in turn correct the Bekenstein-Hawking entropy-area relation. 
		Furthermore, with the reduction in the size of the black hole, quantum
		effects will also start to dominate. Just as the  general relativity can be 
		obtained from thermodynamics in the Jacobson formalism, we propose that
		the quantum fluctuations  to the geometry can be obtained from thermal fluctuations.
	\end{abstract}
	
	\maketitle
	The entropy of a black hole equal to the quarter of the area of its horizon in natural 
	units \cite{1, 1a}. This observation establishes a connection between the 
	thermodynamics and geometry of spacetime.  This entropy associated with a black hole 
	is also the maximum entropy that can be associated with any 
	object of the same volume     \cite{2, 4}. It is interesting to observe that this 
	maximum entropy of a region of space scales with its area and not its volume 
	\cite{4a}. In fact, it is this observation that has motivated the 
	holographic principle \cite{5, 5a}. Even though the holographic principle is 
	a very important principle in physics, it is 
	expected that this holographic principle will get modified near Planck scale due 
	to quantum fluctuations \cite{6, 6a}. This can also be observed from the fact 
	that the relation between the entropy and area of a black hole is expected 
	to get modified due to quantum fluctuations. 
	The leading order correction to the relation between the area and entropy 
	of a black hole is a  logarithmic correction 
	in almost all approach to quantum gravity. 
	Thus, such logarithmic correction have been obtained using   non-perturbative quantum  general
	\cite{1z}, 
	the Cardy formula \cite{card}, matter fields surrounding a black hole \cite{other, other0, other1}, 
	string theory   \cite{solo1, solo2, solo4, solo5}, dilatonic black
	holes \cite{jy}  the partition
	function of a black hole     \cite{bss}, and the 
	generalized uncertainty principle 
	\cite{mi, r1}. 
	Even though the form of the corrections from various different approaches to quantum gravity 
	are logarithmic corrections, the coefficient of such logarithmic correction   is different for 
	all these approaches to quantum quantum. 
	
	It may be noted that such   logarithmic corrections can also be obtained by considering 
	the effects of  thermal fluctuations on the entropy of a black hole  \cite{fl, Das, Landau}.
	Now it is known that in 
	the Jacobson formalism, spacetime emerges from thermodynamics
	\cite{Jacobson}, in that General Relativity can be deduced from the
	Bekenstein-Hawking entropy-area relation combined with the first law of 
	thermodynamics. Thus, the correction to the Bekenstein-Hawking entropy-area relation
	would generate corrections to the structure of spacetime. Furthermore, as the black hole 
	becomes smaller due to hawking radiation, its 
	temperature would increases, and this in turn would increase the 
	contribution coming from the thermal  corrections. However, 
	as the black hole becomes smaller, the effect of quantum fluctuations would also increase. 
	Thus, in this paper, we propose that both these effects are related, due to the relation 
	between the geometry and the black hole  thermodynamics. Furthermore, as Jacobson formalism 
	explicitly relies on this connection, we will use the Jacobson formalism to 
	obtain quantum corrections to the metric by analyzing the correction produce by thermal 
	fluctuations to the thermodynamics of a black hole. 
	
	Thus, we start from the Jacobson formalism, and in this formalism the thermodynamics 
	relation 
	$\delta Q = T dS$ is used to obtain the geometry of spacetime. 
	This is because, it is possible to express the   $Q$ in terms of the
	energy-momentum tensor $T_{ab}$, and use Hawking-Bekenstein relation to relate  
	$S$ to the event horizon area $A$. Thus, we obtain  a geometrical quantity which can
	be expressed in terms of Riemann tensor $R_{ab}$, and so a relation between   
	$T_{ab}$ and $R_{ab}$ is obtained, and this can be demonstrated to be the  Einstein 
	field equations.
	
	More precisely, and following the conventions of \cite{Jacobson}, 
	for any point $p$, one can choose a two-surface element $\cal P$ to
	which orthogonal boosts are generated by a Killing field $\chi^a$ such 
	that the temperature $T$ is taken as the Unruh temperature \cite{Urunuh}
	defined by $T =\hbar \kappa/2\pi$ where $\kappa$ represents the 
	acceleration of the Killing orbit, and the heat flow is then defined 
	by the boost-energy current $T_{ab}\chi^a$.
	As for  the area, we consider a local Rindler horizon through $p$ generated 
	by $\chi^a$ whose future points to the energy carried by matter.
	The past-pointing heat flux through $\cal P$, beyond which lies the horizon
	denoted by $\cal H$,
	\begin{equation}
	\delta Q=\int_{\cal H}T_{ab} \chi^a d\Sigma ^b, \label{1J}
	\end{equation}
	where $d\Sigma ^a = k^a d\lambda d\cal A$ with $k^a$ a tangent vector to the horizon, 
	$\lambda$ an affine parameter vanishing at $\cal P$ with negative values to
	the past of $\cal P$, and $d\cal A$ is the area
	element. Thus, it is possible to write 
	\begin{equation}
	\delta Q=-\kappa\int_{\cal H}\lambda T_{ab} k^a k^b d\lambda d\cal A, \label{2J}
	\end{equation}
	As the entropy $S$ is assumed to be proportional to the horizon area, so  $dS =\eta
	\delta\cal A$. Denoting the expansion of the horizon generated 
	by $\theta$,  we obtain 
	\begin{equation}
	\delta {\cal A}=-\kappa\int_{\cal H}\theta d\lambda d\cal A, \label{3J}
	\end{equation}
	In order now to obtain the  Einstein equations, it suffices to neglect near $\cal P$, the 
	shear $\sigma^2$ and the expansion $\theta$ terms, which vanish at $\cal P$ by a
	suitable choice of the local Rindler horizon,  in the
	Raychaudhuri equation
	\begin{equation}
	\frac{d\theta}{d\lambda}=-\frac{1}{2}\theta^2-\sigma^2-R_{ab}k^ak^b. \label{4J}
	\end{equation}
	Thus, by  integrate this equation, we  find $\theta = -\lambda R_{ab}k^ak^b$, 
	and  Eq. (\ref{3J}) can be expressed as  
	\begin{equation}
	\delta {\cal A}=-\int_{\cal H}\lambda R_{ab} k^a k^b d\lambda d\cal A, \label{5J}
	\end{equation}
	Comparing Eqs. (\ref{2J}) and (\ref{5J}), we observe that 
	$\delta Q = TdS = \left(\hbar\kappa/2\pi\right)\eta\delta\cal A$
	holds provided $T_{ab}k^ak^b =  \left(\hbar\eta/2\pi\right) R_{ab}k^ak^b$
	for all null $k^a$, which leads to $\left(2\pi/\hbar\eta\right)T_{ab} = R_{ab} +fg_{ab}$
	for some function $f$.
	Energy and momentum conservation, combined with contracted Bianchi identities leads to 
	$f = -R/2+\Lambda$ for some constant
	$\Lambda$, and thus we get Einstein equations
	\begin{equation}
	R_{ab}-\frac{1}{2}+\Lambda g_{ab}=\frac{2\pi}{\hbar\eta}T_{ab}
	\end{equation} 
	We would like to point out that  the proportionality 
	constant $\eta$ between the entropy and the area is related to Newton's constant as 
	$G = \left(4\hbar\eta\right)^{-1}$, and hence to Planck length, but the cosmological
	constant $\Lambda$ cannot be related to  other  constants, and thus remains a free parameter 
	even in the Jacobson formalism.
	
	We would like to apply the former approach to determine the quantum corrections on the 
	black hole geometry due to thermal fluctuations.
	We will consider a BTZ black hole as an example, but the formalism developed 
	here can be applied to any black hole geometry. We first observe that in the 
	Jacobson formalism,  the field equations near horizon of a BTZ black hole 
	can be expressed as a thermodynamical 
	identity,  $dE=TdS+P_rdA$, where $E = M$ is the mass of BTZ black hole, $dA$
	is the change in the
	area of the black hole horizon when the horizon is displaced infinitesimally small,
	$P_r$ is the radial
	pressure provided by the source of Einstein equations. It may be noted that 
	since we have $2+1$ dimensional black hole,
	its volume is in fact the area it encloses.
	So, the term ~ $P_rdA$ actually  corresponds to $ PdV$,
	from the general fist law. Furthermore, the  
	pressure $P_r$ is well-defined for
	BTZ black holes, since they are embedded in AdS spacetime.
	This terms has occurred in various previous works on such black holes 
	\cite{Ma:2015llh,Dolan:2012jh,Kubiznak:2014zwa,Kubiznak:2016qmn}.
	
	Now we will analyze the corrections to the entropy of a BTZ black hole due to 
	thermal fluctuations \cite{fl, Das}. 
	Such  thermal fluctuations have been  analyzed as perturbations around the equilibrium, and this 
	has been done by considering the 
	system very close to the equilibrium. So, the approximation  used in this approach is 
	valid as long
	as the correction due to the thermal fluctuations are small compared to the original quantity, i.e.,
	as long as $ \Delta S_0 / S_0 = (S- S_0)/ S_0 << 1$, where $S$ is the corrected entropy
	and $S_0$ is the original entropy of the system.  
	Thus, the ratio of the corrections to the original quantity should be small, and  so the 
	temperature should  not be  large enough to
	produce very large thermal fluctuations \cite{Landau}. 
	
	Now for canonical ensemble with partition function  \cite{fl, Das}, 
	\begin{equation}
	Z = \int_0^\infty  dE\rho (E)\exp(-\beta E),
	\end{equation}
	the density of states  for a system can be written as
	\begin{eqnarray}
	\rho (E) = \frac{1}{2 \pi i} \int^{\beta_0+ i\infty}_{\beta_0 - i\infty}d \beta\exp[S(\beta)],
	\end{eqnarray}
	where $
	S = \beta  E + \ln Z. $
	It may be noted that usually the  entropy is measured around the equilibrium 
	temperature $\beta_0$, and all thermal fluctuations are
	neglected.
	This is done by making the identification $T = \beta^{-1}$.
	However, it is possible to consider the
	thermal fluctuations, and expand the
	entropy  $S(\beta)$ around the equilibrium temperature $\beta_0$ \cite{fl, Das},
	\begin{equation}
	S = S_0 + \frac{1}{2}(\beta - \beta_0)^2
	\left(\frac{\partial^2 S(\beta)}{\partial \beta^2 }\right)_{\beta = \beta_0},
	\label{a1}
	\end{equation}
	where $\beta$ is a temperature close to the equilibrium temperature $\beta_0$, and
	\begin{eqnarray}
	S_0 = \left[S\left(\beta\right)\right]\left. \right|_{\beta= \beta_0}, && S_0^{\prime\prime}
	= \left[\frac{\partial^2 S\left(\beta\right)}{ \partial \beta^2}\right]
	\left. \right|_{\beta = \beta_0}.
	\end{eqnarray}
	Now  density of states can be expressed as
	\begin{eqnarray}
	\rho (E) &=& \frac{\exp(S_0)}{2 \pi i}\int^{\beta_0+ i\infty}_{\beta_0 - i\infty}d\beta
	\nonumber \\  &&
	\times
	\exp \left( \frac{1}{2}(\beta-\beta_0)^2 \left(\frac{\partial^2 S(\beta)}{\partial \beta^2 }
	\right)_{\beta = \beta_0}   \right).
	\end{eqnarray}
	Furthermore, by  a  change of variables, we obtain
	\begin{equation}
	\rho(E) = \frac{\exp(S_{0})}{\sqrt{2\pi}} \left[\left(\frac{\partial^2
		S(\beta)}{\partial \beta^2 }\right)_{\beta = \beta_0}\right]^{- 1/2}.
	\end{equation}
	Thus, it is possible to express $S$ as
	\begin{equation}\label{C1}
	S = S_0 - \frac{\ln  S_0^{\prime\prime}}{2},
	\end{equation}
	It is also possible to express
	the   second derivative of the entropy in  terms of  fluctuations of the energy, and so
	the   corrected entropy can be written as \cite{Das}
	\begin{equation}\label{C2}
	S = S_0 -\frac{1}{2} \ln\left(S_0T^{2}\right).
	\end{equation}
	So, the  thermal fluctuations decrease entropy of the BTZ black hole. 
	It may be noted that black holes have negative heat capacity, so as the temperature  of a black hole increases, its  entropy decreases. 
	This unusual behaviour of black holes  even  occurs, when  thermal fluctuations are neglected.  
	Furthermore, the behavior of original \cite{Carlip:2014pma},  and the correct entropy \cite{Das}, 
	is well known, and this is not the aim of this paper. The main 
	aim of this paper, is to use this corrected entropy to obtain quantum corrections using the Jacobson formalism. 
	
	Let us consider a non-rotating BTZ
	black hole in three-dimensions with metric \cite{banados}
	\begin{equation}
	ds^2=-\left(\frac{r^2}{l^2}-8G_3M\right)
	dt^2+\left(\frac{r^2}{l^2}-8G_3M\right)^{-1}dr^2+r^2d\theta^2 \label{1}
	\end{equation}
	Its Bekenstein-Hawking entropy and Hawking temperature are given by
	\begin{equation}
	S_0=\frac{2\pi r}{4G_3} \label{2}
	\end{equation}
	\begin{equation}
	T=\frac{r}{2\pi l^2}=\left[\frac{G_3}{\pi^2l^2}\right]S_0 \label{3}
	\end{equation}
	where $r =\sqrt{8G_3M}l$ is horizon radius ($G_3 = 3$-dimensional Newton's constant),
	$M$ being mass of
	the black hole and $l$ is related to the cosmological constant by: $\Lambda=-1/l^2$.
	Now, let us suppose the corrected metric has the form
	\begin{equation}
	ds'^2=-F\left(r\right)dt^2+\left(F\left(r\right)\right)^{-1}dr^2+r^2d\theta^2, \label{1'}
	\end{equation}
	where $F\left(r\right)$ is a function of $\left(r\right)$. 
	Here we have assumed that 
	the fluctuations are only $r$ dependent, and they do not have any 
	angular dependence. 
	The justification for this is that any fluctuation that causes a shear 
	to the black hole surface is highly unstable \cite{Edalati:2010hk,McInnes:2013wba}. 
	Since we have  started with a spherically symmetric BTZ black holes, 
	it is a justified to assume that   the  relevant fluctuations 
	are only $r$ dependent.  
	In fact, as  we are considering a black hole near equilibrium,
	and non-rotating, it is necessary for its stability to only consider
	spherically symmetric fluctuations~\cite{Capistrano:2015gma}.

	It also  known that these thermal fluctuations become 
	dominant at high temperatures. As the temperature of a black holes increases 
	as its size decreases, so these thermal fluctuations also increase 
	as the black hole reduces its size. However, as the black hole becomes smaller, 
	quantum effects are also become dominant. As
	these thermal fluctuations scale with the  quantum fluctuations, 
	in the Jacobson formalism (where spacetime emerges 
	from thermodynamics), we can argue that these thermal fluctuations 
	actually produce quantum corrections to this emergent spacetime
	\cite{Jacobson}. It may be noted that the correction to the entropy produced 
	by these thermal fluctuations is a logarithmic correction, and it scales 
	as $ \ln\left(S_0T^{2}\right)$. This 
	also indicates that these thermal corrections in the thermodynamics 
	are related to the quantum corrections to the geometry of spacetime. 
	This is because in almost all approaches to quantum gravity, 
	we obtain a logarithmic correction 
	term as the leading order quantum correction to the black hole entropy 
	\cite{mi, 1z, card, other, other0, other1,solo1, solo2, solo4, solo5, jy, bss,  r1}. 
	Even though a logarithmic correction term is produced in almost all approaches to quantum 
	gravity, this term is propotional to a constant, and that constant
	depends on details of different 
	approaches to quantum gravity. In fact, this constant is
	usually propotional to some new constant 
	in that theory. 
	As this constant depends on the details of the model used, 
	we will use an arbitrary constant   $\alpha$, and 
	define the corrected  microcanonical entropy as 
	\cite{f22, f0, f12, f1, f2,   f44, f4, f5, f6, f7}
	\begin{equation}
	S=S_0+\alpha\ln (S_0T^2)  \label{4}
	\end{equation}
	It may be noted that the    corrected metric should reduces   to
	the original metric, as $\alpha \to 0$, and this occurs if thermal 
	fluctuations are neglected. 
	If we neglect the higher order terms, we write 
	\begin{equation}
	dS=\left(\frac{2\pi}{4G_3}+\frac{\alpha}{r}\right)dr  \label{5}
	\end{equation}
	As $M=\frac{r^2}{8G_3l^2}$, we find that
	\begin{equation}
	dM=\frac{r}{4G_3l^2}dr \label{6}
	\end{equation}
	By substituting Eqs.(\ref{3}),(\ref{5}), and (\ref{6}) in the first thermodynamic law
	\begin{equation}
	dM=TdS+P_rdA
	\end{equation}
	we find
	\begin{equation}
	\frac{r}{4G_3l^2}=\frac{r}{2\pi l^2}\left(\frac{2\pi}{4G_3}+\frac{\alpha}{r}\right)+P_r
	\left(8\pi r\right)
	\end{equation}
	where $A=4\pi r^2$. By simplifying the last equation we find
	\begin{equation}
	-8G_3\pi P_r=-\frac{1}{4l^2}+\frac{1}{4l^2}+\alpha\frac{ G_3}{\pi l^2 \left(2r\right)} \label{7}
	\end{equation}
	Now, Einstein equation is given by
	\begin{equation}
	G_{ab}+\Lambda g_{ab}=-8\pi G_3T_{ab} \label{8'}
	\end{equation}
	The Einstein equation for the metric (\ref{1'}) when evaluated at the horizon radius 
	reads\cite{akbar}
	\begin{equation}
	-8G_3\pi T_{0}^0=\frac{1}{2r}F'\left(r\right)-\frac{1}{l^2} \label{8}
	\end{equation}
	where the prime indicates to the derivative with respect to $r$.
	If we suppose that the static BTZ black hole for the entropy (\ref{4}) is consistent 
	with Jacobson's approach then we can find the corrected metric by
	identifying between  (\ref{7}) and the Eq. (\ref{8}). Then, we get the 
	differential equation
	\begin{equation}
	\frac{1}{2r}F'\left(r\right)-\frac{1}{l^2}=\alpha\frac{G_3}{2\pi l^2r}
	\end{equation}
	By solving it we find
	\begin{equation}
	F\left(r\right)=\frac{r^2}{l^2}+\alpha\frac{ G_3r}{2\pi l^2} +C
	\end{equation}
	where $C$ is the integral constant which is equal to $-8MG_3$ as $F\left(r\right)$ is 
	equal to $f\left(r\right)$ for $\alpha\to 0$.
	Thus, the corrected metric of  (\ref{1}) is given by
	\begin{eqnarray}
	ds'^2 &=& -\left(\frac{r^2}{l^2}+\alpha\frac{ G_3r}{2\pi l^2} -8MG_3\right)dt^2
	\nonumber \\ && +\left(\frac{r^2}{l^2} 
	+\alpha\frac{ G_3r}{2\pi l^2} -8MG_3\right)^{-1}dr^2+r^2d\theta^2
	\label{corrmetric}
	\end{eqnarray}
	Thus, the thermal fluctuations to the thermodynamics   
	can give rise to quantum corrections to the metric in the Jacobson formalism.
	Furthermore, by considering the logarithmic corrections 
	to the thermodynamics, we are only analyzing the first order corrections to the metric.

	It would be interesting to study the corrected thermodynamics from this quantum corrected metric.
	Even though $\alpha$ is a constant, its value depends on the details of the approach. So, here 
	we will analyze the corrected thermodynamics for different values of $\alpha$. 
	The corrected outer and inner horizon's are given by  
	\begin{equation}
	r'_{\pm} = \frac{\sqrt{G_3} \sqrt{\alpha ^2 G_3+128 \pi ^2 l^2 M}\pm\alpha  G_3}{4 \pi }
	\end{equation}
	Now using  the formula for temperature \cite{exactbtz}
	\begin{equation}
	T= \frac{r^2_+-r^2_-}{2 \pi r_+}, 
	\label{temp}
	\end{equation}
	the corrected temperature can be written as 
	\begin{equation}
	T' =\frac{\sqrt{G_3} \left(\alpha ^2 G_3+64 \pi ^2 l^2 M\right)}{2 \pi ^2 \left(\sqrt{\alpha ^2 G+128 \pi ^2 l^2 M}-\alpha  \sqrt{G_3}\right)}
	\end{equation}
	\begin{figure}[h]
		\centering
		\includegraphics[scale=0.5]{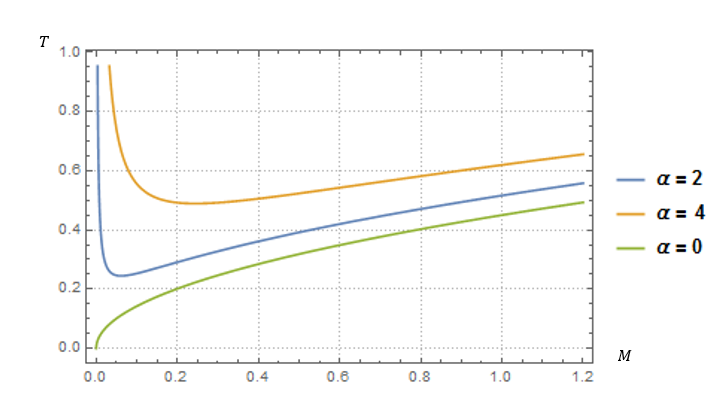}
		\caption{A plot between BTZ temperature and its mass for a fixed $ l=0.5$ in Planckian units. 
			Observe how quantum corrections lead to significant increase of temperature at low masses (small scale). 
			Where quantum fluctuations are expected to be most prominent. }
	\end{figure}
	We may calculate the corrected entropy from the Area law, 
	\begin{equation}
	S' =  4 \pi r'_+ = \sqrt{G_3} \sqrt{\alpha ^2 G_3+128 \pi ^2 l^2 M}-\alpha  G_3
	\end{equation}
	\begin{figure}[h]
		\centering
		\includegraphics[scale=0.5]{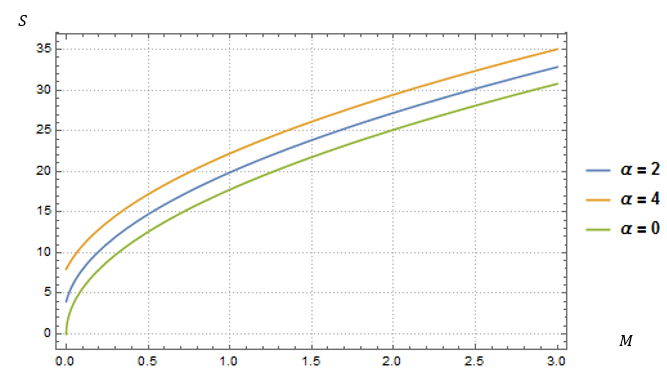}
		\caption{A plot between BTZ entropy and its mass for a fixed $ l=0.5$ in 
			Planckian units. Quantum corrections are mainly relevant for small mass.  }
	\end{figure}
	We may also study the PV criticality from this corrected metric 
	in the extended phase space \cite{wdw}. So,  we can use the definition of thermodynamics pressure as
	\begin{equation}
	P = \frac{T}{v},
	\end{equation}
	with $ v= 2 ( V/ \pi)^{1/2}$, where $V$ is the BTZ black hole volume, given by
	\begin{equation}
	V= 16 \pi r_+^2 = \frac{\left(\alpha  G_3-\sqrt{G_3} \sqrt{\alpha ^2 G_3+128 \pi ^2 l^2 M}\right)^2}{\pi }
	\end{equation}
	Thus, the thermodynamics pressure is  
	\begin{equation}
	P= \frac{\sqrt{G_3} \left(\alpha ^2 G_3+64 \pi ^2 l^2 M\right)}{4 \pi  \left(\zeta-\alpha  \sqrt{G}\right) \sqrt{\left(\alpha  G_3-\sqrt{G_3} \, \zeta\right)^2}}
	\end{equation}
	with $ \zeta =\sqrt{\alpha ^2 G_3+128 \pi ^2 l^2 M}$
	
	Now we are able to calculate the Gibb's free energy $G(T, l, \alpha)$, from the definition
	\begin{equation}
	G = M+PV -TS
	\end{equation}
	The Gibb's free energy determines the critical behaviour of the BTZ black hole,
	if $G>0$ we say the black hole is critical. However, if $G<0$, the black hole is not critical,
	and the saddle points of the $G(M, \alpha) $ Plot indicate the phase transition. 
	The explicit formula of Gibb's free energy is given by
	\begin{eqnarray}
	G &=& -\frac{\alpha ^2 G_3^2}{2 \pi ^2}-32 G_3 l^2 M+M \nonumber \\ &&
	+\frac{\sqrt{G_3} \left(\alpha ^2 G_3+64 \pi ^2 l^2 M\right) \sqrt{\left(\alpha  G_3-\sqrt{G_3} 
			\sqrt{\alpha ^2 G_3+128 \pi ^2 l^2 M}\right)^2}}{4 \pi ^2 \left(\sqrt{\alpha ^2 G_3+128 \pi ^2 l^2 M}-\alpha  \sqrt{G_3}\right)}
	\end{eqnarray}
	As we can see from the plot of $ G(M,\alpha)$, that for this  black hole 
	no critical phenomena exists, and the black hole always remains uncritical (as $ G$ remains less than zero).
	\begin{figure}[h]
		\centering
		\includegraphics[scale=0.6]{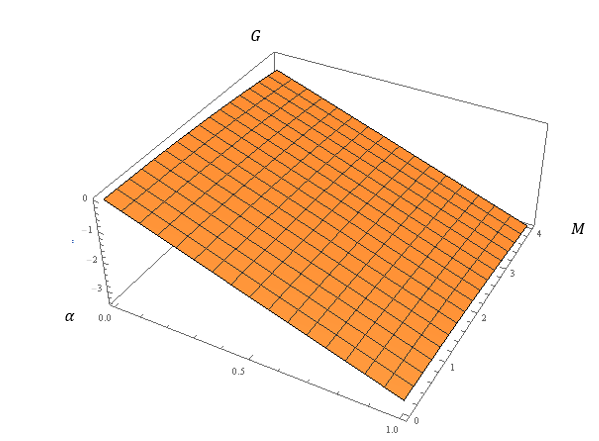}
		\caption{A  3D plot of $G( M, \alpha)$ of quantum corrected  BTZ black hole for a fixed $ l=0.5$. 
			We observe that for different masses and perturbation parameters, the BTZ black hole does not show any critical behaviour, $ G<0$. }
	\end{figure}

	It should be noted as the thermal fluctuations increase we will have to include
	higher order corrections to this perturbative expansion \cite{more}.
	However, near Planck scale the temperature is expected to become sufficiently
	large to break this perturbative expansion, and at this stage the
	system cannot be analyzed as a perturbation around equilibrium temperature.
	This is expected  as general relativity emerges from thermodynamics in the
	Jacobson formalism \cite{Jacobson}, and so we expect that thermal fluctuations
	will occur because of quantum fluctuations.
	It is also known that such logarithmic correction to the entropy are the leading 
	order corrections generated from various different approaches to quantum
	gravity \cite{mi, 1z, card, other, other0, other1,solo1, solo2, solo4, solo5, 
		jy, bss,  r1}. 
	Furthermore, we also expect that as the black hole will become small and
	its  temperature increases, and we have to consider higher order thermal  
	corrections as seen from figure 1.
	\cite{more}, which will correspond to  higher order quantum  corrections
	in the Jacobson formalism \cite{Jacobson}. However, near Planck scale it is expected 
	that the manifold description of the spacetime will breakdown,
	and so we cannot analyze the system  using quantum correction to a classical geometry 
	\cite{manif, manif1}. Similarly, it is expected that the equilibrium description
	of the thermodynamics will breakdown at Planck scale, and we cannot analyze the system
	using thermal fluctuations to the equilibrium thermodynamics.
	
	It would be interesting to analyze this connection between the breaking of the manifold 
	structure of spacetime and the breaking
	of the equilibrium description of the thermodynamics. It might also be interesting to 
	note that by using non-equilibrium thermodynamics, we might
	be able to analyze some purely quantum gravitational states of spacetime near Planck
	scale. The effects of large fluctuations on the behavior of black holes has been studied \cite{larg}, 
	and it would be interesting to analyze such effects using the Jacobson formalism. 
	However, in this paper, we have  only analyze the first order
	corrections to the equilibrium entropy from thermal fluctuations around an equilibrium.
	The important thing to note here is that just as general relativity can emerge
	from thermodynamics in the Jacobson formalism, quantum gravity can emerge from thermal 
	fluctuations to the thermodynamics. Furthermore, just it is possible to
	analyze small quantum fluctuations to the geometry by analyzing small thermal
	fluctuations 
	to the metric. However, at Planck scale, just as we expect manifold description
	to spacetime to breakdown, we also expect the equilibrium discretion to the
	thermodynamics
	to breakdown. We would like to point out that such thermal fluctuations have been studied for various different 
	kind of dynamical black objects \cite{f22, f0, f12, f1, f2,   f44, f4, f5, f6, f7}.  
	It would be interesting to analyze the effects 
	of such thermal fluctuations on the spacetime metric using the formalism developed in this paper. 
	The black hole thermodynamics for time dependent  Vaidya  black holes has also been studied \cite{td01, td02, td04, td05}.
	It would be interesting to analyze the thermal fluctuations for such black holes, and then use the formalism of this paper 
	to obtain the corrected quantum corrected metric for Vaidya black holes.

\end{document}